\documentstyle[12pt]{article}
\catcode`\@=11

\@addtoreset{equation}{section}
\def\theequation{\arabic{section}.\arabic{equation}}

\def\@normalsize{\@setsize\normalsize{15pt}\xiipt\@xiipt
\abovedisplayskip 14pt plus3pt minus3pt%
\belowdisplayskip \abovedisplayskip
\abovedisplayshortskip  \z@ plus3pt%
\belowdisplayshortskip  7pt plus3.5pt minus0pt}

\def\small{\@setsize\small{13.6pt}\xipt\@xipt
\abovedisplayskip 13pt plus3pt minus3pt%
\belowdisplayskip \abovedisplayskip
\abovedisplayshortskip  \z@ plus3pt%
\belowdisplayshortskip  7pt plus3.5pt minus0pt
\def\@listi{\parsep 4.5pt plus 2pt minus 1pt
            \itemsep \parsep
            \topsep 9pt plus 3pt minus 3pt}}

\def\underline#1{\relax\ifmmode\@@underline#1\else
        $\@@underline{\hbox{#1}}$\relax\fi}
\@twosidetrue

\relax

\catcode`@=12

\catcode`\@=11

\def\section{\@startsection{section}{1}{\z@}{3.5ex plus 1ex minus
   .2ex}{2.3ex plus .2ex}{\large\bf}}

\def\thesection{\Roman{section}.}

\def\appendix{\setcounter{section}{0}
        \def\thesection{APPENDIX }
        \def\theequation{\Alph{section}.\arabic{equation}}}

\def\FERMIPUB{}

\def\ps@headings{\def\@oddfoot{}\def\@evenfoot{}
\def\@oddhead{\hbox{}\hfill
        \makebox[.5\textwidth]{\raggedright\ignorespaces --\thepage{}--
        \hfill {\rm FERMILAB--Pub--\FERMIPUB}}}
\def\@evenhead{\@oddhead}
\def\subsectionmark##1{\markboth{##1}{}}
}

\catcode`\@=12

\relax

\def\figcap{\section*{Figure Captions\markboth
        {FIGURECAPTIONS}{FIGURECAPTIONS}}\list
        {Fig. \arabic{enumi}:\hfill}{\settowidth\labelwidth{Fig. 999:}
        \leftmargin\labelwidth
        \advance\leftmargin\labelsep\usecounter{enumi}}}
 \relax
\def\tablecap{\section*{Table Captions\markboth
        {TABLECAPTIONS}{TABLECAPTIONS}}\list
        {Table \arabic{enumi}:\hfill}{\settowidth\labelwidth{Table 999:}
        \leftmargin\labelwidth
        \advance\leftmargin\labelsep\usecounter{enumi}}}
 \relax
\def\reflist{\section*{References\markboth
        {REFLIST}{REFLIST}}\list
        {[\arabic{enumi}]\hfill}{\settowidth\labelwidth{[999]}
        \leftmargin\labelwidth
        \advance\leftmargin\labelsep\usecounter{enumi}}}
 \relax

\catcode`\@=11

\def\FERMIPUB{}

\def\ps@headings{\def\@oddfoot{}\def\@evenfoot{}
\def\@oddhead{\hbox{}\hfill
        \makebox[.5\textwidth]{\raggedright\ignorespaces --\thepage{}--
        \hfill {\rm FERMILAB--Pub--\FERMIPUB}}}
\def\@evenhead{\@oddhead}
\def\subsectionmark##1{\markboth{##1}{}}
}

\ps@headings

\relax
\newskip\humongous \humongous=0pt plus 1000pt minus 1000pt
\def\caja{\mathsurround=0pt}
\def\eqalign#1{\,\vcenter{\openup1\jot \caja
        \ialign{\strut \hfil$\displaystyle{##}$&$
        \displaystyle{{}##}$\hfil\crcr#1\crcr}}\,}
\newif\ifdtup

\def\beq{\begin{equation}}
\def\eeq{\end{equation}}

\def\beqn{\begin{eqnarray}}
\def\eeqn{\end{eqnarray}}
\relax

\def\G2{{\; \rm GeV/}c^2}
\def\G{\; \rm GeV}

\def\dotx{\dotx{\dot\overline{x}}}

\relax

\begin{document}
\hbadness=10000
\begin{titlepage}
\nopagebreak
\begin{flushright}

        {\normalsize KUCP-59\\

        March,~1993}\\
\end{flushright}
\vfill
\begin{center}
{\large \bf Quantum Deformed $su(m|n)$ Algebra and \\
Superconformal Algebra on Quantum Superspace}
\vfill
\renewcommand{\thefootnote}{\fnsymbol{footnote}}
{\bf Tatsuo Kobayashi\footnote{
Fellow of the Japan Society for the Promotion of Science. Work
partially supported by the Grant-in-Aid for Scientific Research from the
Ministry of Education, Science and Culture (\# 030083)}}

       Department of Fundamental Sciences, FIHS, \\
       Kyoto University,~Kyoto 606,~Japan \\
\vfill

\end{center}

\vfill
\nopagebreak
\begin{abstract}
We study a deformed $su(m|n)$ algebra on a quantum superspace.
Some interesting aspects of the deformed algebra are shown.
As an application of the deformed algebra we construct a deformed
superconformal algebra.
{}From the deformed $su(1|4)$ algebra, we derive deformed Lorentz, translation
of Minkowski space, $iso(2,2)$ and its supersymmetric algebras as closed
subalgebras with consistent automorphisms.

\end{abstract}

\vfill
\end{titlepage}
\pagestyle{plain}
\newpage
\voffset = -2.5 cm
\leftline{\large \bf 1. Introduction}
\vspace{0.8 cm}

Recently, quantum groups and quantum algebras [1-4] have attracted much
attention in theoretical physics and mathematics, such as statistical models,
integrable models, conformal field theory and knot theory [5-8].
Also a quantum space has been studied intensively as a non-commutative space
representing the quantum group \cite{Tak,Manin}.
Differential calculus on the quantum space is very intriguing
as an application of the quantum group and useful to show interesting aspects
of the quantum groups [10-15].

Further, quantum Lorentz group and q-deformed Lorentz and Poincar{\' e}
algebras have been constructed on the quantum space [16-19].
The approach was extended to construction of q-deformed conformal and
superconformal algebras on the quantum space \cite{KU2,KU3}.
On the other hand, through the Drinfeld-Jimbo procedure, another q-deformed
Poincar{\' e} and conformal algebras were obtained without use of
the differential calculus of the quantum space \cite{Luki1,Luki2,Luki3}.
As well as the deformed algebras, whole quantum deformed analyses including
representations of the algebrs on the quantum space are very interesting.
Therefore, in this paper we consider deformed algebras on the quantum space,
generalizing the approach of Ref.\cite{SWZ,KU2} to a deformed $su(m|n)$
algebra on the quantum superspace.
As an application, we study deformed superconformal algebra and its closed
subalgebras with some automorphisms.

This paper is organized as follows.
In section two we review the differential calculus on the quantum superspace
and construct a deformed $su(m|n)$ algebra consistent with the space.
For the algebra, a representation on the fermionic space and a relation to the
Drinfeld-Jimbo algebra are shown.
In section three, we introduce the conjugate space and derive a simple
conjugation of the deformed $su(m|n)$ algebra.
In section four we construct a deformed superconformal algebra in
four-dimensional space-time and also give a quantum 6-vector.
{}From the deformed $su(1|4)$ algebra, we derive some closed subalgebrs with
simple consistent automorphisms.
Section five is devoted to conclusions and remarks.
In Appendix A we study decomposition of the quantum differential calculus and
representation of the deformed algebra on a new space and in Appendix B the
deformed $su(1|4)$ algebra is shown explicitly.

\newpage
\leftline{\large \bf 2. Deformed $su(m|n)$ algebra on quantum superspace}
\leftline{\bf 2.1 Quantum superspace}
\vspace{0.8 cm}

The quantum space is a non-commutative space which represents the quantum
groups.
Namely, the quantum group transforms covariantly commutation relations of
the quantum space.
These commutation relations between coordinates $Z^I(x^i,\theta^\alpha)$
$(i=1 \sim m, \alpha = 1 \sim n)$ and derivatives
$\partial_I (\partial_i,\partial_\alpha)$ of the quantum superspace are
obtained in terms of a $\widehat R$-matrix for $GL_q(m|n)$ \cite{WZ,KU1}
as follows,
$$ Z^I Z^J= \widehat R^{IJ}_{\ \ KL}Z^K Z^L, \quad
\partial_I \partial_J= \widehat R^{LK}_{\ \ JI}\partial_K \partial_L,$$
$$ \partial_J Z^I=\delta^I_J +X \widehat R^{IK}_{\ \ JL}Z^L \partial_K,
\eqno(2.1)$$
where $X$ is a deformation parameter.
The $\widehat R$-matrix should satisfy the Yang-Baxter equation and it
involves many deformation parameters in addition to $X$.

In Ref.\cite{KU1}, the quantum group matrices transforming the quantum
superspace has been studied and also it showed conditions on the parameters
for
\renewcommand{\thefootnote}{\fnsymbol{footnote}}
a superdeterminant to be a center \footnote{See in detail Ref.\cite{KU1}}.
Here we consider the differential calculus with one parameter $(X=1/q^2)$,
 making use of the following $\widehat R$-matrix,
$$ \widehat R^{IJ}_{\ \ KL}=\delta^I_L \delta^J_K \{ (( -q^2)^{\widehat I}-
q(-1)^{\widehat I \widehat J})\delta^{IJ}+q(-1)^{\widehat I \widehat J} \}+
(1-q^2)\delta^I_K \delta^J_L \Theta^{JI},
\eqno(2.2)$$
where $\Theta^{JI}=1$ for $I<J$, otherwise vanishes and $\widehat I$ denotes
the grassman parity, i.e., $\widehat I=0$ for $I=i$ and $\widehat I=1$ for
$I=\alpha$.
The above $\widehat R$-matrix leads to a central superdeterminant.
Appearance of $\Theta^{JI}$ implies that ordering of bosonic and fermionic
elements is nontrivial.
Although in this paper we study the ordering where any fermionic element
follows all of bosonic ones, i.e., $\Theta^{\alpha i}=1$, we could discuss
other ordering.

The $\widehat R$-matrix leads to the following commutation relations,
$$ Z^I Z^J={(-1)^{\widehat I \widehat J} \over q }Z^J Z^I,\quad
\partial_I \partial _J=(-1)^{\widehat I \widehat J}q \partial_J \partial_I,
\quad(I<J),$$
$$(\theta^\alpha)^2=(\partial_\alpha)^2=0,\quad
\partial_J Z^I={(-1)^{\widehat I \widehat J} \over q} Z^I \partial_J, \quad
(I\neq J), \eqno(2.3)$$
$$\partial_i x^i=1+q^{-2}x^i \partial_i+(q^{-2}-1)(\sum_{i<j}x^j \partial_j +
\sum_{\alpha}\theta^\alpha \partial_\alpha),$$
$$\partial_\alpha \theta^\alpha=1-\theta^\alpha \partial_\alpha+
(q^{-2}-1)\sum_{\alpha<\beta}\theta^\beta \partial_\beta.$$

In eq.(2.3), the commutation relation between $\partial_I$ and $Z^I$ depends
on other coordinates and derivatives.
In Appendix A, we discuss decomposition of the above differential algebra,
following Ref.\cite{Og} and also representation of the deformed algebra on a
new space is studied.

The $\widehat R$-matrix is decomposed into two projection operators, a
symmetric projector ${\cal S}$ and an antisymmetric one ${\cal A}$ as follows,
$${\cal S}={1 \over 1+q^2}(\widehat R +q^2 {\bf 1}), \quad
{\cal A}={-1 \over 1+q^2}(\widehat R - {\bf 1}).
\eqno(2.4)$$

\vspace{0.8 cm}
\leftline{\bf 2.2 Deformed $su(m|n)$ algebra on quantum superspace}
\vspace{0.8 cm}

Now we consider deformed $su(m|n)$ generators on the quantum superspace.
First of all, we study generators $T^I_{\ I+1}$ and $T^I_{\ I-1}$ associated
with simple roots, which correspond to $Z^I\partial_{I+1}$ and
$Z^I\partial_{I-1}$ in the classical limit ($q \rightarrow 1$).
Following Ref.\cite{SWZ}, we assume that $T^I_{\ I \pm 1}$ acts on $Z^K$ and
$\partial_K$ as follows,
$$ T^I_{\ I\pm 1}Z^K=a(I,K)_\pm Z^K T^I_{\ I\pm 1}+\delta^K_{I\pm 1}Z^I,$$
$$ T^I_{\ I\pm 1}\partial_K=b(I,K)_\pm \partial_K T^I_{\ I\pm 1}+C(I)_\pm
\delta^I_{K}\partial_{I \pm 1}.
\eqno(2.5)$$
The actions of $T^I_{\ I\pm 1}$ (2.5) should be consistent with the
commutation relations of the quantum superspace (2.3).
For example, we calculate $T^I_{\ I+1}(qZ^J Z^K-(-1)^{\widehat J \widehat K}
Z^K Z^J) $ $(J<K)$, so that we derive the following consistency condition,
$$ \delta^J_{I+1}(qZ^I Z^K- (-1)^{\widehat J \widehat K} a(I,K)_+ Z^K Z^I)
+\delta^K_{I+1}(q a(I,J)_+Z^J Z^I-(-1)^{\widehat J \widehat K} Z^I Z^J)=0.
\eqno(2.6)$$
Namely, we obtain $a(I,I)_+=(-1)^{\widehat I\widehat {(I+1)}}/q$ for $I=i$ and
$a(I,J)_+=1$ for $I>J$ or $I+1<J$.
Further, we investigate consistency between (2.5) and the other commutation
relations (2.3), so that we obtain actions of $T^I_{\ I+1}$ as follows,
$$T^i_{\ i +1}x^i=q^{-1}x^iT^i_{\ i +1},\quad
T^i_{\ i +1}\partial_i=q\partial_i T^i_{\ i +1}-q \partial_{i+1},$$
$$T^i_{\ i +1}x^{i+1}=qx^{i+1}T^i_{\ i +1}+x^i,\quad
T^i_{\ i +1}\partial_{i+1}=q^{-1}\partial_{i+1} T^i_{\ i +1},$$
$$T^\alpha_{\ \alpha +1}\theta^\alpha=q\theta^\alpha T^\alpha_{\ \alpha +1},
\quad T^\alpha_{\ \alpha +1}\partial_\alpha=q^{-1}\partial_\alpha
T^\alpha_{\ \alpha +1}-q^{-1} \partial_{\alpha+1},$$
$$T^\alpha_{\ \alpha +1}\theta^{\alpha+1}=q^{-1}\theta^{\alpha+1}
T^\alpha_{\ \alpha +1}+\theta^\alpha,\quad
T^\alpha_{\ \alpha +1}\partial_{\alpha+1}=q\partial_{\alpha+1}
T^\alpha_{\ \alpha +1},\eqno(2.7)$$
$$T^{i=m}_{\ \alpha =1}x^m=q^{-1}x^mT^{i=m}_{\ \alpha =1},\quad
T^{i=m}_{\ \alpha =1}\partial_{i=m}=q\partial_{i=m} T^{i=m}_{\ \alpha =1}
-q \partial_{\alpha =1},$$
$$T^{i= m}_{\ \alpha =1}\theta^1=-q^{-1}\theta^1 T^{i=m}_{\ \alpha =1}+x^m,
\quad T^{i=m}_{\ \alpha =1}\partial_{\alpha =1}=-q\partial_{\alpha =1}
T^{i=m}_{\ \alpha =1}.$$
For the other elements, $T^I_{\ I+1}$ satisfy the classical relations, i.e.,
they commute or anticommute with each other depending on their grassman
parities.

Similarly we can derive actions of $T^{I+1}_{\ I}$ as follows,
$$T^{i+1}_{\ i}x^i=q^{-1}x^iT^{i+1}_{\ i}+x^{i+1},\quad
T^{i+1}_{\ i}\partial_i=q\partial_i T^{i+1}_{\ i},$$
$$T^{i+1}_{\ i}x^{i+1}=qx^{i+1}T^{i+1}_{\ i},\quad
T^{i+1}_{\ i}\partial_{i+1}=q^{-1}\partial_{i+1} T^{i+1}_{\ i}-q^{-1}\partial_i
,$$
$$T^{\alpha +1}_{\ \alpha}\theta^\alpha=q\theta^\alpha
T^{\alpha +1}_{\ \alpha}+\theta^{\alpha+1},
\quad T^{\alpha +1}_{\ \alpha}\partial_\alpha=q^{-1}\partial_\alpha
T^{\alpha +1}_{\ \alpha},$$
$$T^{\alpha +1}_{\ \alpha}\theta^{\alpha+1}=q^{-1}\theta^{\alpha+1}
T^{\alpha +1}_{\ \alpha},\quad
T^{\alpha +1}_{\ \alpha}\partial_{\alpha+1}=q\partial_{\alpha+1}
T^{\alpha +1}_{\ \alpha}-q\partial_\alpha,\eqno(2.8)$$
$$T^{\alpha =1}_{\ i =m}x^m=q^{-1}x^mT^{\alpha =1}_{\ i=m}+\theta^1,\quad
T^{\alpha =1}_{\ i=m}\partial_{i=m}=q\partial_{i=m} T^{\alpha =1}_{\ i=m},$$
$$T^{\alpha =1}_{\ i=m}\theta^1=-q^{-1}\theta^1 T^{\alpha =1}_{\ i=m},\quad
T^{\alpha =1}_{\ i=m}\partial_{\alpha =1}=-q\partial_{\alpha =1}
T^{\alpha =1}_{\ i=m}+q\partial_{i=m}.$$
For the other elements, $T^{I+1}_{\ I}$ satisfy the classical relations.

Next, we define Cartan generators $H_I$ in terms of a commutation relation
between $T^I_{\ I+1}$ and $T^{I+1}_{\ I}$.
In the classical limit, $H_I$ corresponds to $Z^{I+1}\partial_{I+1}
-(-1)^{\widehat I\widehat {(I+1)}}Z^I\partial_I$.
For example, we can obtain actions of $T^{i+1}_{\ i}T^{i}_{\ i+1}$ and
$T^{i}_{\ i+1}T^{i+1}_{\ i}$ making use of (2.7) and (2.8) as follows,
$$T^{i+1}_{\ i}T^{i}_{\ i+1}x^i=q^{-2}x^iT^{i+1}_{\ i}T^{i}_{\ i+1}
+q^{-1}x^{i+1}T^{i}_{\ i+1},$$
$$T^{i+1}_{\ i}T^{i}_{\ i+1}x^{i+1}=q^2x^{i+1}T^{i+1}_{\ i}T^{i}_{\ i+1}
+q^{-1}x^iT^{i+1}_{\ i}+x^{i+1},$$
$$T^{i}_{\ i+1}T^{i+1}_{\ i}x^i=q^{-2}x^iT^{i}_{\ i+1}T^{i+1}_{\ i}
+qx^{i+1}T^i_{\ i+1}+x^i,
\eqno(2.9)$$
$$T^{i}_{\ i+1}T^{i+1}_{\ i}x^{i+1}=q^2x^{i+1}T^{i}_{\ i+1}T^{i+1}_{\ i}
+qx^iT^{i+1}_{\ i}.$$
Following Ref.\cite{SWZ}, we define Cartan generators $H_i$ $(i=1\sim m-1)$ by
linear
combination of $T^{i+1}_{\ i}T^{i}_{\ i+1}$ and $T^{i}_{\ i+1}T^{i+1}_{\ i}$
in order to eliminate the linear terms of $T$ on the right hand side of (2.9)
as follows,
$$ H_i\equiv qT^{i+1}_{\ i}T^{i}_{\ i+1}-q^{-1}T^{i}_{\ i+1}T^{i+1}_{\ i}.
\eqno(2.10)$$

Similarly we define the other Cartan generators $H_0(=H_{i=m})$ and
$H_\alpha$ ($\alpha =1\sim n$) in terms of $T^{i=m}_{\ \alpha =1}$,
$T^{\alpha =1}_{\ i=m}$,$T^{\alpha}_{\ \alpha +1}$ and
$T^{\alpha +1}_{\ \alpha}$ so that linear terms of $T$ disappear in actions
of $H_0$ and $H_\alpha$.
Thus we have
$$ H_\alpha \equiv q^{-1}T^{\alpha +1}_{\ \alpha}T^{\alpha }_{\ \alpha +1}
-qT^{\alpha }_{\ \alpha+1}T^{\alpha +1}_{\ \alpha}, \quad
H_0 \equiv \{ T^{i=m}_{\ \alpha =1}, T^{\alpha =1}_{\ i=m} \},
\eqno(2.11)$$
and actually the Cartan generators act on the quantum superspace as follows,
$$H_ix^i=q^{-2}x^iH_i-q^{-1}x^i, \quad
H_i\partial_i=q^{2}\partial_iH_i+q\partial_i,$$
$$H_ix^{i+1}=q^{2}x^{i+1}H_i+qx^{i+1}, \quad
H_i\partial_{i+1}=q^{-2}\partial_{i+1}H_i-q^{-1}\partial_{i+1},$$
$$H_\alpha \theta^\alpha=q^{2}\theta^\alpha H_\alpha-q\theta^\alpha, \quad
H_\alpha \partial_\alpha=q^{-2}\partial_\alpha H_\alpha+
q^{-1}\partial_\alpha,
\eqno(2.12)$$
$$H_\alpha \theta^{\alpha+1}=q^{-2}\theta^{\alpha+1}H_\alpha+
q^{-1}\theta^{\alpha+1}, \quad
H_\alpha \partial_{\alpha +1}=q^{2}\partial_{\alpha +1}H_\alpha
 -q\partial_{\alpha+1},$$
$$H_0 x^m=q^{-2}x^m H_0+x^m, \quad
H_0 \partial_{i=m}=q^{2}\partial_{i=m} H_0-q^{2}\partial_{i=m},$$
$$H_0 \theta^{1}=q^{-2}\theta^{1}H_0+\theta^{1}, \quad
H_0 \partial_{\alpha =1}=q^{2}\partial_{\alpha =1}H_0
 -q^{2}\partial_{\alpha=1},$$
and for the other elements the generators satisfy the classical algera.
Note that the definition of $H_0$ (2.11) is never deformed.

Finally, we define the other generators $T^I_{\ J}$ $(J\neq I \pm 1)$  in terms
 of commutation relations of $T^I_{\ I\pm 1}$.
For example we define $T^i_{\ i+2}$ as
$$T^i_{\ i+2} \equiv [T^i_{\ i+1},T^{i+1}_{\ i+2}]_h,
\eqno(2.13)$$
where $[A,B]_h \equiv AB-hBA$.
Then we investigate closure of their algebra.
Using (2.7) and (2.13), we can easily calculate actions of
$T^i_{\ i+1}T^i_{\ i+2}$ and $T^i_{\ i+2}T^i_{\ i+1}$ on $x^k$ as follows,
$$[T^i_{\ i+1}T^i_{\ i+2},x^i]_{1/q^2}=[T^i_{\ i+2}T^i_{\ i+1},x^i]_{1/q^2}
=0,$$
$$[T^i_{\ i+1}T^i_{\ i+2},x^{i+1}]_q=x^{i}(T^i_{\ i+2}+q^{-1}(q^{-1}-h)
T^i_{\ i+1}T^{i+1}_{\ i+2}),$$
$$[T^i_{\ i+1}T^i_{\ i+2},x^{i+2}]_q=(q(q-h)x^{i+1}T^i_{\ i+1}
+(q+q^{-1}-h)x^i)T^i_{\ i+1},
\eqno(2.14)$$
$$[T^i_{\ i+2}T^i_{\ i+1},x^{i+1}]_q=x^{i}(q^{-1}T^i_{\ i+2}+q(q^{-1}-h)
T^{i+1}_{\ i+2}T^{i}_{\ i+1}),$$
$$[T^i_{\ i+2}T^i_{\ i+1},x^{i+2}]_q=((q-h)x^{i+1}T^i_{\ i+1}
+x^i)T^i_{\ i+1}.$$
Eq.(2.14) shows that if $h=q$ or $1/q$, $hT^i_{\ i+1}T^i_{\ i+2}$ is identified
 with $T^i_{\ i+2}T^i_{\ i+1}$.
Here we choose $h=1/q$, so as to obtain
$$[T^i_{\ i+1},T^i_{\ i+2}]_q=0.
\eqno(2.15)$$

Similarly we difine the generators $T^I_{\ K}$ as follows,
$$T^I_{\ K} \equiv T^I_{\ J}T^J_{\ K}-q_{\widehat J}T^J_{\ K}T^I_{\ J}, \quad
(I<J<K {\rm \ or \ } I>J>K),
\eqno(2.16)$$
where $q_{\widehat J}=q^{2\widehat J-1}$, i.e., $q_{\widehat J}=1/q$ for $J=i$
and $q_{\widehat J}=q$ for $J=\alpha$.
The definition satisfy the algebraic closure similar to the above.
The generators $T^I_{\ K}$ ($I<K$) act on the quantum superspace as follows,
$$T^I_{\ K}Z^I=q_{\widehat I}Z^I T^I_{\ K}, \quad
T^I_{\ K}Z^J=(-1)^{(\widehat I+ \widehat K)\widehat J}Z^J T^I_{\ K},$$
$$T^I_{\ K}Z^K = (-1)^{(\widehat I+1)\widehat K}(q_{\widehat k})^{-1}Z^K
T^I_{\ K}+Z^I+(-1)^{\widehat I}\lambda\sum_{I<J<K}Z^J T^I_{\ J},$$
$$T^I_{\ K}\partial_J=(-1)^{(\widehat I+ \widehat K)\widehat J}\partial_J
T^I_{\ K}-(-1)^{\widehat I}\lambda \partial_KT^I_{\ J},
\eqno(2.17)$$
$$T^I_{\ K} \partial_I=(q_{\widehat I})^{-1}(\partial_IT^I_{\ K}-\partial_K),
\quad
T^I_{\ K}\partial_K= (-1)^{(\widehat I+1)\widehat K} q_{\widehat k} \partial_K
T^I_{\ K},$$
where $\lambda=q-1/q$ and $I<J<K$ and for the other elements the generators
satisfy the classical relations.
Also the generators $T^K_{\ I}$ ($I<K$) act on the quantum superspace as
follows,
$$T^K_{\ I}Z^I=q_{\widehat I}Z^IT^K_{\ I}+Z^K, \quad
T^K_{\ I}Z^K=(-1)^{(\widehat I+1)\widehat K}(q_{\widehat K})^{-1}Z^KT^K_I,$$
$$T^K_{\ I}Z^J=(-1)^{(\widehat I +\widehat K)\widehat J}Z^JT^K_{\ I}+
(-1)^{\widehat I}\lambda Z^KT^K_{\ I},$$
$$T^K_{\ I}\partial_I=(q_{\widehat I})^{-1}\partial_IT^K_{\ I}, \quad
T^K_{\ I}\partial_J=(-1)^{(\widehat I+\widehat K)\widehat J}\partial_J
T^K_{\ I},$$
$$T^k_{\ i}\partial_k=q^{-1}\partial_kT^k_{\ i}-q^{2(i-k)+1}\partial_i
-\lambda \sum _{i<j<k}q^{2(j-k)}\partial_jT^j_{\ i},
\eqno(2.18)$$
$$T^\gamma_{\ \alpha}\partial_\gamma=q\partial_\gamma T^\gamma_{\ \alpha}-
q^{2(\gamma -\alpha )-1}\partial_\alpha +\lambda \sum _{\alpha <\beta <\gamma }
q^{2(\gamma -\beta )}\partial_\beta T^\beta_{\ \alpha},$$
$$T^\beta_{\ i}\partial_\beta=-q\partial_\beta T^\beta_{\ i}+
q^{2(\beta +i-m)-1}\partial_i +\lambda (\sum_{i<j}q^{2(\beta +j-m-1)}
\partial_j T^j_{\ i}-\sum_{\alpha <\beta}q^{2(\beta -\alpha)}\partial_\alpha
T^\alpha_{\ i}),$$
where $I<J<K$, $i<j<k$ and $\alpha < \beta < \gamma$, and for the other
elements the generators satisfy the classical relations.

{}From (2.12), (2.17) and (2.18), we can derive commutation relations of
generators through the calculation similar to (2.14) and (2.15).
Among the whole algebra, commutation relations of $T^I_{\ I\pm 1}$ and $H_I$
are obtained as
$$[T^{I-1}_{\ I},T^{I+1}_{\ I}]_{q_{\widehat I}}=
[T^{I}_{\ I+1},T^{I}_{\ I-1}]_{q_{\widehat I}}=[T^I_{\ J},T^K_{\ L}]=0,
\quad (I,J \neq K,L),$$
$$[H_I,T^I_{\ I+1}]_{(q_{\widehat I})^4}=-(q_{\widehat I})^2(q+q^{-1})
T^I_{\ I+1}, \quad
[H_I,T^{I+1}_{\ I}]_{1/(q_{\widehat I})^4}=(q_{\widehat I})^{-2}(q+q^{-1})
T^{I+1}_{\ I},$$
$$[H_I,T^J_{\ J+1}]_{1/(q_{\widehat I})^2}=(q_{\widehat I})^{-1}T^J_{\ J+1},
\quad [H_I,T^{J+1}_{\ J}]_{(q_{\widehat I})^2}=-q_{\widehat I}T^{J+1}_{\ J},
\quad (I=J\pm 1), $$
$$[H_0,T^{i=m}_{\ \alpha =1}]=[H_0,T^{\alpha =1}_{\ i=m}]=0
\eqno(2.19)$$
$$[H_0,T^{i=m-1}_{\ i=m}]_{q^2}=-q^2T^{i=m-1}_{\ i=m}, \quad
[H_0,T^{i=m}_{\ i=m-1}]_{1/q^2}=T^{i=m}_{\ i=m-1},$$
$$[H_0,T^{\alpha =1}_{\ \alpha =2}]_{1/q^2}=T^{\alpha =1}_{\ \alpha =2},
\quad
[H_0,T^{\alpha =2}_{\ \alpha =1}]_{q^2}=-q^2T^{\alpha =2}_{\ \alpha =1},$$
where $H_I$ do not involve $H_0$.
Further, it is easily shown that the Cartan generators commute with each other
and $T^i_{\ \alpha}$ and $T^\alpha_{\ i}$ are nilpotent, i.e.,
$(T^i_{\ \alpha})^2=(T^\alpha_{\ i})^2=0$.
The other commutation relations of the whole algebra are obtained by use of
(2.19) or the calculation similar to (2.14) and (2.15), through boresomely
long calculations.
For a concrete example, all commutation relations of the deformed $su(1|4)$
algebra is found explicitly in Appendix B, while the whole deformed $su(4)$
algebra has been shown in Ref.\cite{KU2}.
It is remarkable that the deformed $su(\ell)$ algebra on the quantum bosonic
space is represented in the same way as on the quantum fermionic space,
 except replacing $q$ by $1/q$.
This replacement is not essential.

\vspace{0.8 cm}
\leftline{\bf 2.3 Representation of deformed $su(2)$ algebra on quantum
fermionic space }
\vspace{0.8 cm}

In this and the following subsections, two interesting topics about the above
deformed $su(\ell)$ algebra are discussed.
Here we study the deformed $su(2)$ algebra on the two-dimensional quantum
fermionic space, whose coordinates ($\theta^1,\theta^2$) and derivatives
($\partial_1,\partial_2$) satisfy the following commutation relation,
$$q\theta^1 \theta^2=-\theta^2 \theta^1, \quad
\partial_1 \partial_2=-q\partial_2 \partial_1, $$
$$q \partial _\alpha \theta^\beta=-\theta^\beta \partial_\alpha, \quad
(\alpha \neq \beta), \eqno(2.20)$$
$$\partial_1 \theta^1=1-\theta^1 \partial_1+(q^{-2}-1)\theta^2 \partial_2,
\quad \partial_2 \theta^2=1-\theta^2 \partial_2. $$

Suppose that we define $\widehat T^1_{\ 2}$ and $\widehat T^2_{\ 1}$ like the
classical generators using the quantum elements as follows,
$$\widehat T^1_{\ 2}\equiv \theta^1 \partial_2, \quad
\widehat T^2_{\ 1} \equiv \theta^2 \partial_1.
\eqno(2.21)$$
Then they satisfy the following relations,
$$\widehat T^1_{\ 2} \theta^2=q^{-1}\theta^2\widehat T^1_{\ 2}+\theta^1,\quad
\widehat T^2_{\ 1} \theta^1=q\theta^1\widehat T^2_{\ 1}+\theta^2,$$
$$\widehat T^1_{\ 2} \partial_1=q^{-1}\partial_1\widehat T^1_{\ 2}
-q^{-1}\partial_2,\quad
\widehat T^2_{\ 1} \partial_2=q\partial_2\widehat T^2_{\ 1}
-q\partial_1.
\eqno(2.22)$$
The above relations are nothing but the actions of the genarators
$T^1_{\ 2}$ and $T^2_{\ 1}$ of the deformed $su(2)$ algebra on the quantum
fermionic space (2.7) and (2.8).

Next, in the similar way to (2.8) we define $\widehat H$ using
$\widehat T^1_{\ 2}$ and $\widehat T^2_{\ 1}$ as follows,
$$\widehat H \equiv q^{-1} \widehat T^2_{\ 1}\widehat T^1_{\ 2}
-q\widehat T^1_{\ 2}\widehat T^2_{\ 1}=
q^{-1}\theta^2 \partial_2-q\theta^1\partial_1.
\eqno(2.23)$$
The generator $\widehat H$ satisfies the following relations,
$$\widehat H \theta^1=q^2\theta^1 \widehat H-q\theta^1,\quad
\widehat H \theta^2=q^{-2}\theta^2 \widehat H+q^{-1}\theta^1,$$
$$\widehat H \partial_1=q^{-2}\partial_1 \widehat H+q^{-1}\partial_1,\quad
\widehat H \partial_2=q^2\partial_2 \widehat H-q\partial_1.
\eqno(2.24)$$
The above relations are also exactly same as the actions of $H_1$ (2.12) of
the deformed $su(2)$ algebra.
Therefore the deformed $su(2)$ algebra is completely represented in terms of
 the coordinates and derivatives of the quantum fermionic space (2.21) and
(2.23).
Unfortunately, for the bosonic case we can not represent the deformed $su(2)$
algebra by the quantum bosonic coordinates and derivatives.

\newpage
\vspace{0.8 cm}
\leftline{\bf 2.4 Map to Drinfeld-Jimbo basis}
\vspace{0.8 cm}

In this subsection, we discuss a map between the above deformed algebra and
Drinfeld-Jimbo's algebra.
The deformed $su(\ell)$ algebra on the bosonic and the fermionic quantum
spaces involves the following deformed $su(2)$ algebra as a closed subalgebra,
$$[H_\alpha,T^\alpha_{\ \alpha +1}]_{q^4}=-q^2(q+q^{-1})T^\alpha_{\ \alpha +1},
\quad [H_\alpha,T^{\alpha +1}_{\ \alpha}]_{1/q^4}=q^{-2}(q+q^{-1})
T^{\alpha +1}_{\ \alpha},$$
$$ [T^{\alpha +1}_{\ \alpha},T^\alpha_{\ \alpha+1}]_{q^2}=qH_\alpha,
\eqno(2.25)$$
where if we replace $q$ by $1/q$, we obtain the deformed $su(2)$ algebra on
the quantum bosonic space, which consists of $T^i_{i +1}$, $T^{i+1}_{i}$
and $H_i$.
This algebra (2.25) is identified with a deformed $su(2)$ algebra by
Woronowicz, up to a normalization factor \cite{Woro}.

On the other hand, Drinfeld and Jimbo constructed the following deformed
$su(2)$ algebra with a deformation parameter $q'$ as follows,
$$[J_0,J_+]=J_+, \quad [J_-,J_0]=J_-$$
$$[J_+,J_-]={1 \over 2}[2J_0]_{q'},
\eqno(2.26)$$
where $[A]_q=(q^A-q^{-A})/(q-q^{-1})$.

In Ref.\cite{CZ,CFZ}, a map between the Woronowicz's and the Drinfeld-Jimbo's
$su(2)$ algebras was discussed.
Thus, changing a normalization factor on the map of Ref.\cite{CZ,CFZ},
we can easily derive a map from the Drinfeld-Jimbo algebra (2.26) with $q'=q$
to (2.25) as follows,
$$ H_\alpha=\lambda^{-1}(1-q^{-4J_0}), \quad
T^\alpha_{\ \alpha_+1}=\sqrt 2 q^{-J_0}J_-, \quad
T^{\alpha +1}_{\ \alpha}=\sqrt 2 q^{-J_0}J_+.
\eqno(2.27)$$
Under the similar map, the deformed $su(2)$ algebra on the quantum bosonic
space is related with the Drinfeld-Jimbo's algebra with $q'=1/q$.
The above map could be generalized to the deformed $su(\ell)$, because
we can relate the Cartan generators and the generators associated with the
simple roots of two basis through (2.27).

The map (2.27) leads to actions of $J_0$ and $J_\pm$ on the quantum space.
For example, in the case of (2.27) with $\alpha =1$ we have
$$q^{-4J_0}\theta^1=q^2\theta^1 q^{-4J_0}, \quad
q^{-4J_0}\theta^2=q^{-2}\theta^2q^{-4J_0},$$
$$q^{-4J_0}\partial_1=q^{-2}\partial_1 q^{-4J_0}, \quad
q^{-4J_0}\partial_2=q^2\partial_2q^{-4J_0},$$
$$J_+\theta^1=\sqrt q\theta^1J_+ +\sqrt {q \over 2}\theta^2 q^{J_0}, \quad
J_+ \theta^2={1 \over \sqrt q} \theta^2 J_+,$$
$$J_+\partial_1={1 \over \sqrt q}\partial_1 J_+, \quad
J_+ \partial_2={\sqrt q}\partial_2 J_+ - \sqrt {q^3 \over 2}\partial_2 q^{J_0},
\eqno(2.28)$$
$$J_-\theta^1=\sqrt q\theta^1J_-, \quad
J_- \theta^2={1 \over \sqrt q} \theta^2 J_- +{1 \over \sqrt {2q}}\theta^1
q^{J_0},$$
$$J_-\partial_1={1 \over \sqrt q}\partial_1 J_- -{ 1 \over \sqrt {2q}}
\partial_2 q^{J_0}, \quad
J_- \partial_2={\sqrt q}\partial_2 J_- .$$
Similarly we can derive actions of the Drinfeld-Jimbo generators on the
quantum bosonic space.

In Ref.\cite{CZ}, it was shown maps to other deformed $su(2)$ algebra, e.g.,
the two Witten's algebras \cite{Wit} and the Fairlie's algebra \cite{Fair}.
Through the precedure, the deformed algebra obtained here could be related to
other algebra and we could derive relations between the quantum space and other
 algebra.

\vspace{0.8 cm}
\leftline{\large \bf 3. Conjugation}
\vspace{0.8 cm}

In this section, we introduce another quantum space $\overline Z_I$ conjugate
to $Z^I$.
We set up commutation relations of $\overline Z_I$ as follows,
$$ \overline Z_L \overline Z_K = \widehat R^{IJ}_{\ KL} \overline Z_J
\overline Z_I, \quad
\overline Z_K Z^I = \widehat R^{IJ}_{\ KL} Z^L \overline Z_J.
\eqno(3.1)$$
The latter relation is explicitly written as
$$\overline Z_J Z^I=(-1)^{\widehat I \widehat J}qZ^I \overline Z_J,$$
$$\overline x_i x^i = x^i \overline x_i+(1-q^2)(\sum_{i<j}x^j \overline x_j
+\sum_\alpha \theta^\alpha \overline \theta_\alpha),
\eqno(3.2)$$
$$\overline \theta_\alpha \theta^\alpha=-q^2 \theta^2 \overline \theta_\alpha +
(1-q^2)\sum_{\alpha < \beta} \theta^\beta \overline \theta_\beta.$$
These commutation relations have a center $\sum_I Z^I \overline Z_I$.
Further, we assume that $\overline Z_I$ has the same commutation relations
with the generators as $\partial_I$.
For example, we have
$$ T^{i+1}_{\ i}\overline x_{i+1}=q^{-1}\overline x_{i+1}T^{i+1}_{\ k}
-q^{-1}\overline x_i.
\eqno(3.3)$$

Now we relate $\overline Z_I$ and $Z^I$ by a conjugation, where
$\overline Z_I$ must be proportional to the complex conjugate of $Z^I$.
The conjugation should be consistent with the commutation relations (2.3) and
(3.1).
We can find two types of consistent conjugations, depending on the value of
$q$.
In the case with real $q$, we can consistently relate $Z^I$ and
$\overline Z_I$ as follows,
$$ \overline Z_I=g_I(Z^I)^*,
\eqno(3.4)$$
where $g_I$'s are diagonal elements of a metric for $su(m_1,m_2|n_1,n_2)$ and
$*$ implies the complex conjugate.
In this case, when taking the conjugation, we have to reverse the order of
elements, i.e., $\overline {ab}=\overline b \overline a$.

On the other hand, when $|q|=1$, the following conjugation is consistent,
$$\overline x_i =ig_iKq^{m-i+1}(x^i)^*, \quad
\overline \theta_\alpha = g_\alpha q^\alpha K (\theta^\alpha)^*,
\eqno(3.5)$$
where $K$ is an arbitrary phase factor.
In this case, when taking the conjugation, we do not reverse the order of
elements, i.e., $\overline {ab}= \overline a \overline b$.

Now we consider conjugation of the generators.
Here, we restrict ourselves the case with real $q$.
We could extend the following approach to the case where $|q|=1$.
Actually, in Ref.\cite{KU2,KU3}, the conjugation of the deformed $su(4)$ and
$su(1|4)$
algebras was disscussed in the case where $|q|=1$.
For example we take conjugation of the relation between $T^i_{\ i+1}$ and
$x^{i+1}$ (2.7), so that we obtain
$$\overline x_{i+1} \overline {T^i_{\ i+1}}=q \overline {T^i_{\ i+1}}
\overline x_{i+1} +g_i g_{i+1} \overline x_i.
\eqno(3.6)$$
Through comparison with (3.3), eq.(3.6) leads to the following conjugation
relation,
$$\overline T^i_{\ i+1} = g_i g_{i+1} T^{i+1}_{\ i}.
\eqno(3.7)$$
Similarly we can derive the conjugation relations of the other generators as
follows,
$$\overline {T^I_{\ J}} = g_I g_J T^J_{\ I}, \quad \overline {H_I} =H_I, \quad
\overline H_0=H_0.
\eqno(3.8)$$

\newpage
\leftline{\large \bf 4. Deformed superconformal algebra on quantum
space}
\vspace{0.8 cm}

In this section, we study a deformed superconformal algebra as a very
interesting application of the deformed $su(m|n)$ algebra on the quantum
superspace.
The N=1 superconformal algebra in the four-dimensional space-time is
represented by $su(1|2,2)$, whose deformed algebra is found in Appendix B.
Therefore, we need the quantum space which consists of one bosonic and four
fermionic elements to represent the algebra.
In this section, the bosonic coordinate and derivative are denoted by $x^0$ and
 $\partial_0$, while $\theta^\alpha$ and $\partial_\alpha$ ($\alpha=1 \sim 4$)
implies the fermionic elements as the previous sections.
Further, here we choose the $su(2,2)$ metric as $g=(1,1,-1,-1)$.

\vspace{0.8 cm}
\leftline{\bf 4.1 Quantum 6-vector}
\vspace{0.8 cm}

In the classical limit, the $su(2,2)$ algebra is isomorphic to $so(4,2)$.
In Ref.\cite{JO,KU2}, the isomorphism has been generalized to the quantum
case.
In the papers, a quantum 6-vector was discussed in terms of bi-spinors.
It is an important representaion as well as the spinor representation,
which corresponds to $\theta^\alpha$.
In this subsection, we consider a repsentation of the deformed superconformal
algebra $su_q(1|2,2)$, which includes the quantum 6-vector.

Now we construct the quatum 6-vector in terms of a tensor product of two
quantum spinors.
For that purpose, we introduce another spinor $\eta^\alpha$
($\alpha = 1\sim 4$) and $y^0$.
The new quantum superspace $y^0$ and $\eta^\alpha$ represent the deformed
$su(1|2,2)$ algebra in the same way as $x^0$ and $\theta^\alpha$.
We assume that they satisfy the following commutation relations,
$${Z_1}^I{Z_2}^J=\widehat R^{IJ}_{\ KL}{Z_2}^K{Z_1}^L,
\eqno(4.1)$$
where $Z_1\equiv (x^0,\theta^\alpha)$ and $Z_2\equiv (y^0,\eta^\alpha)$.
Eq.(4.1) is explicity written as follows,
$$x^0y^0=y^0x^0,\quad x^0\eta^\alpha=q\eta^\alpha x^0+(1-q^2)y^0\eta^\alpha,$$
$$\theta^\alpha y^0=qy^0\theta^\alpha, \quad
\theta^\alpha \eta^\alpha =-q^2\eta^\alpha \theta^\alpha,
\eqno(4.2)$$
$$\theta^\alpha \eta^\beta=-q\eta^\beta \theta^\alpha +(1-q^2)\eta^\alpha
\theta^\beta, \quad \theta^\beta \eta^\alpha = -q\eta^\alpha \theta^\beta,
\quad (\alpha < \beta).$$

We use the projection operators ${\cal S}$ and ${\cal A}$ in order to decompose
 the tensor product $Z_1Z_2$ into irreducible representations of the deformed
 $su(1|2,2)$ algebra.
Although in Ref.\cite{KU2} the antisymmetric projector ${\cal A}$ was used to
decompose the tensor product of two bosonic quantum spinors into the quantum
6-vector, the symmetric projector ${\cal S}$ is available to derive another
 quantum 6-vector from a tensor product of two fermionic quantum space.
Actually, from ${\cal S}^{IJ}_{\ KL}{Z_1}^K{Z_2}^L$ we obtain eleven
independent elements as follows,
$$ S^{\alpha \beta}=\theta^\alpha \eta^\beta-q\theta^\beta \eta^\alpha,
\quad (\alpha < \beta ),$$
$$ S^{0\alpha} =x^0\eta^\alpha+q\theta^\alpha y^0, \quad S^{00}=x^0y^0,
\eqno(4.3)$$
where $S^{\alpha \beta}$ is the quantum 6-vector.

{}From (2.12), (2.17) and (2.18), we can derive actions of the deformed
$su(1|2,2)$ generators on $S$.
The Cartan generators act on $S$ as follows,
$$[H_\alpha,S^{\alpha \beta}]_{q^2}=-qS^{\alpha \beta},\quad
[H_\alpha,S^{\alpha +1 \ \beta}]_{1/q^2}=q^{-1}S^{\alpha +1 \ \beta},$$
$$[H_\alpha,S^{I \ \alpha+1}]_{1/q^2}=q^{-1}S^{I \ \alpha +1}, \quad
[H_\alpha,S^{I \ \alpha}]_{q^2}=-qS^{I \ \alpha},
\eqno(4.4)$$
$$[H_0,S^{0 J}]_{1/q^4}=(1+q^{-4})S^{0 J}, \quad
[H_0,S^{J \alpha}]_{1/q^2}=S^{J \alpha},$$
where $I<\alpha$, $\alpha +1<\beta$ and $J=0,1$.
Similarly, we obtain actions of $T^K_{\ L}$ $(K<L)$ as follows,
$$[T^\alpha_{\ \gamma},S^{\alpha \gamma}]=-q\lambda\sum_{\alpha <\beta< \gamma}
S^{\alpha \beta}T^\alpha_{\ \beta},$$
$$\{T^0_{\ \beta},S^{0\beta} \}_{1/q^2}=(q+q^{-1})S^{00}+{\lambda \over q}
\sum_{\alpha <\beta}S^{0\alpha}T^0_{\ \alpha},$$
$$[T^\alpha_{\ \beta},S^{I \alpha}]_q=[T^I_{\sigma},
S^{I\rho}]_{(-1)^{\widehat I}\widehat q(I)}=0, \quad (\sigma \neq \rho),$$
$$[T^\alpha_{\ \gamma },S^{I\gamma}]_{1/q}=S^{I\alpha }-\lambda \sum_{\alpha <
\beta < \gamma }S^{I \beta }T^{\alpha}_{\ \beta},
\eqno(4.5)$$
$$[T^I_{\ \beta},S^{\beta \gamma}]_{1/q}=S^{I\gamma}-\lambda \sum_{I < \alpha
\beta }S^{\alpha \beta }T^{I}_{\ \alpha},$$
$$[T^I_{\ \gamma},S^{\alpha \gamma}]_{1/q}=-qS^{I\alpha}-\lambda \sum_{\alpha <
\beta < \gamma}S^{\alpha \beta }T^{I}_{\ \beta}+q\lambda \sum_{I<\rho <\alpha }
S^{\rho \alpha } T^0_{\ \rho},$$
where $I<\alpha < \beta < \gamma$.
Further, we have actions of $T^L_{\ K}$ ($K<L$) as,
$$[T^\beta_{\ \alpha},S^{\alpha \gamma}]_q=\{T^\beta_{\ 0},S^{0 \gamma}
\}_{1/q}=S^{\beta \gamma},$$
$$[T^\alpha_{\ I},S^{\alpha \beta}]_{1/q}=\{T^\alpha_{\ 0},S^{0 \alpha }
\}_{1/q}=[T^\alpha_{\ I},S^{J \alpha}]_{1/q}=0,$$
$$[T^\beta_{\ \alpha},S^{I \alpha}]_{1/q}=S^{I\beta},$$
$$[T^\gamma_{\ \alpha},S^{\alpha \beta}]_q=-qS^{\beta \gamma}
-q\lambda S^{\alpha \gamma}T^\beta_{\ \alpha},
\eqno(4.6)$$
$$\{ T^\beta_{\ 0},S^{ 0 \alpha} \}_{1/q}=-qS^{\alpha \beta}+{ \lambda \over q}
 S^{0 \beta}T^\alpha_{\ 0},$$
$$[T^\gamma_{\ \alpha},S^{I\beta}]=-\lambda S^{I \gamma}T^\beta_{\ \alpha},
\quad
[T^\beta_{\ I},S^{\alpha \gamma}]=-\lambda S^{\beta \gamma}T^\alpha_{\ I},$$
$$[T^\gamma_{\ I},S^{\alpha \beta}]=-\lambda S^{\alpha \gamma}T^\beta_{\ I}+
q\lambda S^{\alpha \beta}T^\alpha_{\ I},$$
where $I\neq J$ and $I,J<\alpha <\beta < \gamma$.
Note that $S^{0 \alpha}$ has the same relations with the deformed $su(2,2)$
generators as the quantum 4-spinor $\theta^\alpha$.

\vspace{0.8 cm}
\leftline{\bf 4.2 Deformed superconformal algebra}
\vspace{0.8 cm}

In this section we assign the deformed $su(1|4)$ generators to the physical
 superconformal generators (See e.g. Ref.\cite{van}).
The assignment should be consistent with the conjugation (3.8).
First of all, it is convenient to assign the deformed $su(2,2)$ generators to
a deformed $so(4,2)$ generators $M_{\mu \nu}$ ($\mu,\nu=0 \sim 5$) as follows,
$$\eqalign{
&M_{12}=\textstyle{-1\over 2}(H_1+H_3), \quad
M_{23}=\textstyle{-1\over 2}({T^1}_2+{T^2}_1+{T^3}_4+{T^4}_3), \cr
&M_{31}=\textstyle{i\over 2}({T^1}_2-{T^2}_1+{T^3}_4-{T^4}_3), \quad
M_{01}=\textstyle{-1\over 2}({T^1}_4-{T^4}_1+{T^2}_3-{T^3}_2), \cr
&M_{02}=\textstyle{i\over 2}({T^1}_4+{T^4}_1-{T^2}_3-{T^3}_2),\quad
M_{03}=\textstyle{-1\over 2}(-{T^1}_3+{T^3}_1+{T^2}_4-{T^4}_2), \cr
&M_{40}=\textstyle{-i\over 2}({T^1}_3+{T^3}_1+{T^2}_4+{T^4}_2), \quad
M_{41}=\textstyle{-1\over 2}({T^1}_2+{T^2}_1-{T^3}_4-{T^4}_3) \cr
&M_{42}=\textstyle{i\over 2}({T^1}_2-{T^2}_1-{T^3}_4+{T^4}_3), \quad
M_{43}=\textstyle{-1\over 2}(H_1-H_3)\cr
&M_{50}=\textstyle{-1\over 2}(H_1+2H_2+H_3), \quad
M_{51}=\textstyle{i\over 2}({T^2}_3+{T^3}_2+{T^1}_4+{T^4}_1) \cr
&M_{52}=\textstyle{-1\over 2}({T^2}_3-{T^3}_2-{T^1}_4+{T^4}_1), \quad
M_{53}=\textstyle{i\over 2}(-{T^1}_3-{T^3}_1+{T^2}_4+{T^4}_2) \cr
&M_{45}=\textstyle{-1\over 2}({T^1}_3-{T^3}_1+{T^2}_4-{T^4}_2). \cr
}\eqno(4.7)$$
In the classical limit, the generators $M_{\mu \nu}$ satisfy the following
algebra,
$$[M_{\mu \nu},M_{\rho\sigma}]=-ig_{\mu \rho}M_{\nu \sigma}+ig_{\nu \rho}
M_{\mu \sigma}+ig_{\mu \sigma}M_{\nu \rho}-ig_{\nu \sigma}M_{\mu \rho},
\eqno(4.8)$$
where the $so(4,2)$ metric is choosen $g_{\mu \nu}= {\rm diag}\
(1,-1,-1,-1,-1,1)$.
Under the conjugation (3.8), $M_{\mu \nu}$ is \lq  real ', i.e., they satisfy
$$\overline M_{\mu \nu} =M_{\mu \nu}.
\eqno(4.9)$$
Next we \lq  compactify ' two-dimensional space, e.g., fourth and fifth
dimensions.
Namely we choose generators for translation $P_\mu$ ($\mu =0 \sim 3$),
conformal boost $K_{\mu}$ and dilatation $D$ as,
$$P_{\mu}=M_{4\mu}+M_{5\mu},\quad K_{\mu}=M_{5\mu}-M_{4\mu}, \quad D=M_{45}.
\eqno(4.10)$$
Further we assign the supercharges $Q_\alpha$, $\overline Q_{\dot \alpha}$,
$S^{\alpha}$, $\overline{S}^{\dot \alpha}$ as follows,
$$\eqalign{
&Q_1= \textstyle{\sqrt{2}}({T^0}_1-i{T^0}_3), \quad
\overline{Q}_1= \textstyle{\sqrt{2}}({T^1}_0-i{T^3}_0), \cr
&Q_2= \textstyle{\sqrt{2}}(-{T^0}_2+i{T^0}_4), \quad
\overline{Q}_2= \textstyle{\sqrt{2}}(-{T^2}_0+i{T^4}_0), \cr
&\overline{S}^1= \textstyle{\sqrt{2}}({T^0}_1+i{T^0}_3), \quad
S^1= \textstyle{\sqrt{2}}({T^1}_0+i{T^3}_0,) \cr
&\overline{S}^2= \textstyle{\sqrt{2}}(-{T^0}_2-i{T^0}_4), \quad
S^2= \textstyle{\sqrt{2}}(-{T^2}_0-i{T^4}_0), \cr
}\eqno(4.11)$$
so that in the classical limit they satisfy
$$\{ Q_\alpha,\overline
Q_{\dot \alpha} \}=2(\sigma^\mu)_{\alpha \dot \beta}P_\mu, \quad
\{ \overline S^{\dot \alpha},S^{\beta} \}=2(\overline \sigma^\mu)^{\dot
\alpha \beta}K_\mu.
\eqno(4.12)$$
At last we choose the U(1) charge as,
$$A=-\textstyle{1\over 4}(4H_0+3H_1+2H_2+H_3).
\eqno(4.13)$$
{}From the deformed $su(1|4)$ algebra, we can read off the deformed
superconformal algebra of the generators defined in the above.
However, as the algebra in the basis is very complicated, it is convenient to
 represent the algebra in the $T$-$H$ basis.
It is interesting that the supercharges are not longer nilpotent.
For example, $Q_1$ satisfy the following relations,
$$(Q_1)^2=2i(q-1)T^0_{\ 3}T^0_{\ 1}, \quad (Q_1)^3=0.
\eqno(4.14)$$
The other supercharges satisfy similar relations.

It is very important that in this basis either the deformed Poincar\'e or the
super-Poincar\'e algebrs is not a closed subalgebra of the deformed
superconformal algebra.
Therefore, we need some contraction procedure \cite{Cele} to obtain a closed
Poincar\'e or super-Poincar\'e algebra from the above basis
\cite{Luki1,Luki2,Luki3,KU2,KU3}.

We restricted ourselves deformation of the N=1 superconformal algebra on
four-dimensional space-time in the above.
It is easy to extend the above approach to extended superconformal algebra on
four-dimensional and two-dimensional space, i.e., $su(N|2,2)$ and $su(N|1,1)$.

\vspace{0.8 cm}
\leftline{\bf 4.3 Closed subalgebra of deformed $su(1|4)$ algebra with
automorphism}
\vspace{0.8 cm}

In this subsection, we investigate closed subalgebras with some simple
automorphisms, i.e., other conjugations besides (3.8).
In Ref.\cite{KU2}. it was shown that $H_\alpha$, $T^\alpha_{\ \alpha+1}$ and
$T^{\alpha +1}_{\ \alpha}$ ($\alpha =1,3$) compose a closed subalgebra (2.25)
for a deformed Lorentz algebra with the following consistent automorphism,
$$H_1 \leftrightarrow H_3, \quad T^1_{\ 2} \leftrightarrow T^4_{\ 3}, \quad
 T^2_{\ 1} \leftrightarrow T^3_{\ 4}.
\eqno(4.15)$$

Next we consider another closed subalgebra consists of $T^1_{\ 3}$,$T^1_{\ 4}$,
$T^2_{\ 3}$ and $T^2_{\ 4}$.
They satisfy the following relations,
$$ [T^1_{\ 3},T^2_{\ 4}]=\lambda [T^1_{\ 4},T^2_{\ 3}], \quad
 [T^1_{\ 4},T^2_{\ 3}]=0
,$$
$$[T^2_{\ 3},T^1_{\ 3}]_q=[T^1_{\ 4},T^1_{\ 3}]_q=[T^2_{\ 4},T^2_{\ 3}]_q
=[T^2_{\ 4},T^1_{\ 4}]_q=0.
\eqno(4.16)$$

In the case where $|q|=1$, this algebra has the following automorphism,
$$T^1_{\ 3} \rightarrow T^1_{\ 3},\quad T^2_{\ 4} \rightarrow T^2_{\ 4}, \quad
T^1_{\ 4} \leftrightarrow T^2_{\ 3}.
\eqno(4.17)$$
When we take the above \lq conjugation ' with $|q|=1$ , we have to reverse
the order of elements.
The fact is opposite to the conjugation discussed in section three.
For this closed subalgebra with the conjugation (4.17), we can find
translation operators $\tilde P_\mu$ ($\mu=0 \sim 4$) with the Minkowski
metric as follows,
$$
\left(
\begin{array}{cc}
\tilde P_0+\tilde P_3 & \tilde P_1+i\tilde P_2\\
\tilde P_1-i\tilde P_2 & \tilde P_0-\tilde P_3\\
\end{array}
\right)\equiv
\left(
\begin{array}{cc}
T^1_{\ 3} & T^1_{\ 4}\\
T^2_{\ 3} & T^3_{\ 4}\\
\end{array}
\right).
\eqno(4.18)$$
It is easily shown that $\tilde P_{\mu}$ is real under the conjugation (4.17).
Further the above algebra has a center $C$ as follows,
$$C=qT^1_{\ 3}T^2_{\ 4}-T^1_{\ 4}T^2_{\ 3}.
\eqno(4.19)$$

Although supercharges associated with the translation operators $\tilde P_\mu$
correspond to linear combinations of $T^1_{\ 0}$,$T^2_{\ 0}$,$T^0_{\ 3}$ and
$T^0_{\ 4}$, unfortunately the supercharges and $\tilde P_{\mu}$ do not
compose a closed algebra and their commutation relations involuve $H_0$,
${H_1}$ and $T^2_{\ 1}$.
We are very interested in a closed Poincar\'e algebra including
$\tilde P_\mu$.
However, we can not find a Lorentz algebra consistent with the conjugation
(4.17).

Instead of a deformed Poincar\'e algebra, we can derive an interesting deformed
 inhomogeneous $iso(2,2)$ algebra with a metric
$\widehat g_{\mu \nu}={\rm diag}( 1,-1,1,-1)$ composed by translation
$T^1_{\ 3}$,$T^1_{\ 4}$,$T^2_{\ 3}$,$T^2_{\ 4}$ and rotation
$T^\alpha_{\ \alpha+1}$,$T^{\alpha +1}_{\ \alpha}$,$H_\alpha$ ($\alpha =1,3$).
They satisfy a closed subalgebra, which is read off Appendix B.
We define a physical basis of translation and rotation as follows,
$$
\left(
\begin{array}{cc}
\widehat P_0+\widehat P_3 & \widehat P_1+\widehat P_2\\
\widehat P_1-\widehat P_2 & \widehat P_0-\widehat P_3\\
\end{array}
\right)\equiv i
\left(
\begin{array}{cc}
T^1_{\ 3} & T^1_{\ 4}\\
T^2_{\ 3} & T^3_{\ 4}\\
\end{array}
\right).
$$
$$\widehat M_{03}={i \over 2}(H_1-H_3),\quad \widehat M_{21}={i \over 2}(H_1+
H_3),
\eqno(4.20)$$
$$\widehat M_{23}={i \over 2}(T^1_{\ 2}+T^2_{\ 1}+T^3_{\ 4}+T^4_{\ 3}),
\quad \widehat M_{13}={i \over 2}(T^1_{\ 2}-T^2_{\ 1}+T^3_{\ 4}-T^4_{\ 3}),$$
$$\widehat M_{02}={i \over 2}(-T^1_{\ 2}+T^2_{\ 1}+T^3_{\ 4}-T^4_{\ 3}),
\quad \widehat M_{01}={i \over 2}(-T^1_{\ 2}-T^2_{\ 1}+T^3_{\ 4}+T^4_{\ 3}).$$
In the case where $|q|=1$, this deformed algebra of $\widehat P$,
$\widehat M$ has the following automorphism,
$$ H_\alpha \rightarrow -H_\alpha,\quad T^\alpha_{\ \alpha+1} \rightarrow
-qT^\alpha_{\ \alpha+1}, \quad T^{\alpha +1}_{\ \alpha} \rightarrow
{-1 \over q}T^{\alpha +1}_{\ \alpha}, \quad
T^\beta_{\ \gamma} \rightarrow -T^\beta_{\ \gamma},
\eqno(4.21)$$
where $\alpha=1,3,\ \beta =1,2,\ \gamma=3,4$.
When we take the above cojugation, we also reverse the order of elements.
Under the conjugation (4.21) $\widehat P_\mu$ and $\widehat M_{\mu \nu}$ are
real.
We can take away the minus sign of the last equation of (4.21) if we put off
$i$ from the definition of $\widehat P_{\mu}$ (4.20).
Further it is remarkable that $C$ (4.19) is still the center in the deformed
$iso(2,2)$ algebra.

At last we consider supersymmetrization of the above deformed $iso(2,2)$
algebra.
Supercharges associated with $\widehat P$ correspond to $T^1_{\ 0}$,
$T^2_{\ 0}$,$T^0_{\ 3}$,$T^0_{\ 4}$.
The elements need $H_0$ so that they have closed commutation relations
with the generators of the deformed $iso(2,2)$ generators.
The superalgebra has the following automorphism consistent with (4.21),
$$T^1_{\ 0} \rightarrow fT^1_{\ 0},\quad T^2_{\ 0} \rightarrow
{f \over q^2}T^2_{\ 0},\quad
H_0 \rightarrow {-1 \over q^2}H_0, \quad
T^0 _{\ \alpha} \rightarrow {-1 \over q^3 f}T^0_{\ \alpha},
\eqno(4.22)$$
where $\alpha =3,4$ and $f$ is an arbitrary factor.
In the last equation of (4.22), the minus sign is due to the sign of the last
equation (4.21).
So we can take away both signs simultaniously.
Suppose that we define physical superchareges as follows,
$$\widehat Q_1 \equiv iT^1_{\ 0},\quad \widehat Q_2 \equiv iq^{-1} T^2_, \quad
\overline {\widehat Q_1} \equiv -q^{-3/2}T^0_{\ 3}, \quad
\overline {\widehat Q_2} \equiv -q^{-3/2}T^0_{\ 4},
\eqno(4.23)$$
where we choose $f=1$.
They satisfy classicaly commutation relations similar to (4.12).
Note that $\widehat Q_\alpha$ and
$\overline {\widehat Q_\alpha}$ are \lq real' themselves under the conjugation
(4.22),i.e., $\overline {\widehat Q_\alpha}$ does not implies the conjugate of
$\widehat Q_\alpha$ and they are independent operators.
The approach to the above superalgebra could be extended.

\vspace{0.8 cm}
\leftline{\large \bf 5. Conclusion}
\vspace{0.8 cm}

We have studied here the deformed $su(m|n)$ algebra on the quantum superspace.
Some interesting aspects of the algebra has been shown and we have constructed
the deformed superconformal algebra as an application of the deformed $su(m|n)$
 algebra.
The quantum 6-vector is also obtained from the tensor product of the fermionic
quantum 4-spinors.
Further we have discussed the closed subalgebras of the deformed $su(1|4)$
algebra with the consistent automorphisms, which include the deformed Lorentz,
translation of Minkowski space, $iso(2,2)$ and its supersymmetric algebras.

It is very important to apply the above approach to deformed $so$ and $sp$
algebras.
The deformed $su(m|n)$ algebra obtained here includes lots of interesting
algebra, e.g., the extended 2-dim and 4-dim superconformal algebra to be
studied and the deformed $iso(2,2)$ algebra, whose classical space is
interesting, e.g., for the O(2) string (See e.g. Ref.\cite{o2}).
It might be possible to derive a deformed Poincar\'e algebra with the correct
metric and reality, from a large algebra in the similar way to the procedure
to find the closed subalgebra with the simple automorphism in section 4.3.

\vspace{0.8 cm}
\leftline{\large \bf Acknowledgement}
\vspace{0.8 cm}

The author would like to thank T.~Uematsu for numerous valiable discussions
and reading the manuscript and P.~P.~Kulish and R.~Sasaki for helpful
discussions.
He also thanks S.~Matsuda and H.~Aoyama for encouragements.

\newpage
\leftline{\large \bf Appendix A}
\vspace{0.8 cm}

Here, we study decomposition of the differential algebra (2.3) following Ref.
\cite{Og}, where decomposition of the bosonic differential algebra was disscued
 through \lq \lq renormalization " of the quantum coordinates and derivatives.
We define
$$ \mu_i \equiv \partial_i x^i-x^i\partial_i=1+(q^{-2}-1)(\sum_{i\leq j}x^j
\partial_j +\sum_\alpha \theta^\alpha \partial_\alpha),$$
$$ \mu_\alpha \equiv \partial_\alpha \theta^\alpha+\theta^\alpha
\partial_\alpha =1+(q^{-2}-1)(\sum_{\alpha <\beta} \theta^\beta \partial_\beta)
.\eqno(A.1)$$
They satisfy the following commutation relations,
$$[\mu_I,Z^J]_{1/q^2}=[\mu_I,\partial_J]_{q^2}=[\mu_i,x^i]_{1/q^2}=
[\mu_i,\partial_i]_{q^2}=0$$
$$[\mu_J,Z^I]=[\mu_J,\partial_I]=[\mu_\alpha,\theta^\alpha]=[\mu_\alpha,
\partial_\alpha]=0,
\eqno(A.2)$$
where $I<J$, and $\mu_I$ commute with each other.

Suppose we define new coordinates $\widehat Z^I$ ($X^i,\Theta^\alpha$) and
derivaties $D_\alpha$ as follows,
$$X^i=\mu_i^{-1/2}x^i, \quad \Theta^\alpha=\mu_\alpha^{-1/2}\theta^\alpha,
\quad D_I=\mu_I^{-1/2}\partial_I.
\eqno(A.3)$$
Then the new coordinates and derivatives satisfy simple relation as
$$D_iX^i=1+q^2X^i\partial_i,\quad D_\alpha\Theta^\alpha=1-\Theta D_\alpha,$$
$$[\widehat Z^I,\widehat Z^J]=[\widehat Z^I,D_J]=[D_I,D_J]=0,
\eqno(A.4)$$
It is remarkable that the fermionic elements $\Theta^\alpha$ and $D_\alpha$
satisfy colmpletely the classical algebra.
Therfore, we can derive the quantum fermionic coordinates $\theta^\alpha$
and derivatives $\partial_\alpha$ from the classical elements $\Theta^\alpha$
 and $D_\alpha$ as follows,
$$\theta^\alpha=\sqrt {\mu_\alpha} \Theta^\alpha, \quad \partial_\alpha =
\sqrt {\mu_\alpha} D_\alpha,$$
$$\mu_\alpha=\prod_{\alpha < \beta }(1+(q^{-2}-1)\Theta^\beta D_\beta).
\eqno(A.5)$$
The similar relation has been found for the fermionic q-oscilators in Ref.
\cite{CKL}.

Next we study actions of the generators on the new sapce.
For example, we consider the case of the two-dimensional bosonic quantum
space with coordinates $x^i$ $(i=1,2)$ and derivatives $\partial_i$.
By definition (A.1), we can easily obtain
$$[T^1_{\ 2},\mu_1]=[T^2_{\ 1},\mu_1]=0,$$
$$[T^1_{\ 2},\mu_2]=(q^{-2}-1)x^1\partial_2=(q^{-1}-1)(\sqrt {\mu_2} x^1 D_2
+x^1D_2\sqrt {\mu_2}),
\eqno(A.6)$$
$$[T^2_{\ 1},\mu_2]=(1-q^{-2})x^2\partial_1=q^{-1}(q^{-1}-1)(\sqrt {\mu_2}X^2
\partial_1+X^2\partial_1\sqrt {\mu_2}).$$
Eq.(A.6) leads to the following relations,
$$[T^1_{\ 2},\sqrt {\mu_2}]=(q^{-1}-1)(\mu_2)^{-1/2}x^1\partial_2, \quad
[T^2_{\ 1},\sqrt {\mu_2}]=q^{-1}(q^{-1}-1)(\mu_2)^{-1/2}x^2\partial_1.
\eqno(A.7)$$
Therefore, we obtain actions of $T^1_{\ 2}$ and $T^2_{\ 1}$ on the renormalized
 coordinates $X^1$ and $X^2$ as follows,
$$T^1_{\ 2}X^1=-q^{-1}X^1T^1_{\ 2},\quad
T^1_{\ 2}X^2=qX^2T^1_{\ 2}+q\mu X^1+q(q-1)\mu X^1X^2D_2,$$
$$T^2_{\ 1}X^1=q^{-1}X^1T^2_{\ 1}+\mu^{-1}X^2,\quad
T^2_{\ 1}X^2=qX^2T^1_{\ 2}+(1-q^{-1})\mu X^2X^2D_1,
\eqno(A.8)$$
where $\mu=\sqrt {\mu_1/\mu_2}$.
This result is rather complicated than (2.7) and (2.8).
We could obtain actions of $T^1_{\ 2}$ and $T^2_{\ 1}$ on the renormarized
fermionic quantum sapce $\Theta^\alpha$, which are also complicated.

\newpage
\leftline{\large \bf Appendix B}
\vspace{0.8 cm}
Here, the whole deformed $su(1|4)$ algebra is explicitly shown.
$$[T^I_{\ J},T^J_{\ K}]_q=T^I_{\ K}, \quad (I<J<K {\rm or } I>J>K).$$
$$[T^{\alpha +1}_{\ \alpha}, T^\alpha_{\ \alpha +1}]_{q^2}=qH_\alpha,
\quad \{ T^0_{\ 1},T^1_{\ 0}\}=H_0,$$
$$[H_\alpha,T^\alpha_{\ \alpha +1}]_{q^4}=-q^2(q+q^{-1})T^\alpha _{\ \alpha +1}
,\quad
[H_\alpha,T^{\alpha +1}_{\ \alpha}]_{1/q^4}={q+q^{-1} \over q^2}T^{\alpha +1}
_{\ \alpha},$$
$$ [H_\alpha ,T^\alpha_{\ J}]_{q^2}=-qT^\alpha_{\ J}, \quad
[H_{\alpha},T^{\alpha +1}_{\ J}]_{1/q^2}=q^{-1}T^{\alpha +1}_{\ J},
\quad (\alpha +1<J {\rm or } J<\alpha),$$
$$ [H_\alpha ,T^J_{\ \alpha}]_{1/q^2}=q^{-1}T^J_{\ \alpha}, \quad
[H_{\alpha},T^J_{\ \alpha +1}]_{q^2}=-qT^J_{\ \alpha +1},
\quad (\alpha +1<J {\rm or } J<\alpha),$$
$$ [H_0,T^J_{\ \rho}]_{1/q^2}=T^J_{\ \rho}, \quad
[H_0,T^\rho_{\ J}]_{q^2}=-q^2T^\rho_{\ J}, \quad (J=0,1\ \rho >1),$$
$$[T^\alpha_{\ \beta},T^I_{\ \beta}]_q=[T^\beta_{\ I},T^\beta_{\ \alpha}]_q=
[T^I_{\ \alpha},T^\beta_{\ \alpha}]_q=0,$$
$$[T^I_{\ \beta},T^I_{\ \alpha}]_{(-1)^{\widehat I}\widehat q(I)}=
[T^\alpha_{\ I},T^\beta_{\ I}]_{(-1)^{\widehat I}q}=
[T^\alpha_{\ \beta},T^\alpha_{\ I}]_q=0,$$
$$[T^\alpha_{\ \gamma},T^I_{\ \beta}]=\lambda T^I_{\ \gamma}T^\alpha_{\ \beta},
\quad [T^\beta_{\ I},T^\gamma_{\ \alpha}]=\lambda T^\gamma_{\ I}
T^\beta_{\ \alpha},$$
$$[T^{\alpha +1}_{\ \alpha},T^\alpha_{\ \rho}]_q=T^{\alpha+1}_{\ \rho}
-\lambda T^{\alpha+1}_{\ \rho}H_\alpha, \quad (\alpha+1<\rho),$$
$$[T^\rho_{\ \alpha},T^\alpha_{\ \alpha+1}]_q=q^2T^\rho_{\ \alpha+1}
-\lambda q^2T^\rho_{\ \alpha+1}H_\alpha, \quad (\alpha +1<\rho),$$
$$[T^\sigma_{\ \gamma},T^\gamma_{\ \rho}]_{1/q}=T^\sigma_{\ \rho}, \quad
(\sigma<\rho,\rho=\gamma-1) {\rm \ or \ } (\rho<\sigma,\sigma=\gamma-1),$$
$$[T^\sigma_{\ 4},T^4_{\ \rho}]_{q^{-1}}=T^\sigma_{\ \rho}-\lambda T^3_{\ \rho}
T^\sigma_{\ 3},\quad ((\sigma,\rho)=(1,2) {\rm \ or \ } (2,1)),$$
$$\eqalign{[T^1_{\ \sigma},T^\rho_{\ 1}]_{q^{-1}}= & -q^{\rho-\sigma+2}
T^\rho_{\ \sigma}+\lambda T^\rho_{\ 2}T^2_{\ \sigma}+\lambda q^{\rho-\sigma+2}
T^\rho_{\ \sigma}(H_1+H_2) -\lambda^2T^\rho_{\ 2}T^2_{\ \sigma}H_1 \cr
& -\lambda^2q^{\rho-\sigma+2}T^\rho_{\ \sigma}H_1H_2,
\hskip 2cm ((\sigma,\rho)=(3,4) {\rm \ or \ } (4,3)),}$$
$$\eqalign{[T^\alpha_{\ \alpha+2},T^{\alpha+2}_{\ \alpha}]_{q^{-2}}= & -q^{-1}
H_\alpha-qH_{\alpha+1}-\lambda q^{-1}T^{\alpha+1}_{\ \alpha}
T^\alpha_{\ \alpha +1}+\lambda q^{-1}T^{\alpha+2}_{\ \alpha+1}
T^{\alpha+1}_{\ \alpha+2} \cr
& +q\lambda H_\alpha H_{\alpha+1}-\lambda^2 q^{-1}T^{\alpha+1}_{\  \alpha+2}
T^{\alpha+2}_{\ \alpha+1}H_{\alpha},}$$
$$\eqalign{[T^1_{\ 4},T^4_{\ 1}]_{q^{-2}}=-q(2-q^{-2})H_1-q^{-1}(2-q^2)H_2-
q(2-q^2)H_3-q^{-1}\lambda T^4_{\ 2}T^2_{\ 4}+\lambda q^{-1}T^1_{\ 2}T^2_{\ 1}
\cr
-q^{-1}\lambda T^3_{\ 1}T^1_{\ 3}+q^{-1}(2-q^2)\lambda T^4_{\ 3}T^3_{\ 4}+
(1+q^{-2})\lambda T^4_{\ 2}T^2_{\ 4}H_1 -\lambda qH_1H_2 -\lambda q^3 H_1 H_3
\cr
+\lambda q^{-1} (2-q^{-2})(2-q^2) H_2 H_3-(q+q^{-1})
-(q+q^{-1})\lambda^2T^3_{\ 2}T^2_{\ 3}H_1+\lambda^2 T^3_{\ 2} T^2_{\ 3} \cr
-q^{-3}(2-q^2)\lambda^2T^4_{\ 3}T^3_{\ 4}H_2+q\lambda^2 T^4_{\ 3}T^3_{\ 4}H_1
-q\lambda^3T^4_{\ 3} T^3_{\ 4}H_1H_2+q^2\lambda^3H_1H_2H_3,}$$
$$[T^0_{\ \gamma},T^\gamma_{\ \alpha}]_{1/q}=q^{2(\gamma -\alpha -1)}
T^0_{\ \alpha}-\lambda \sum_{\alpha < \beta < \gamma}q^{2(\gamma -\beta )-1}
T^0_{\ \beta}T^\beta_{\ \alpha},$$
$$[T^\gamma_{\ 0},T^\alpha_{\ \gamma}]_q=-q^{2(\gamma -\alpha )-1}
T^\alpha_{\ 0}+\lambda \sum_{\alpha < \beta < \gamma}q^{2(\gamma -\beta )}
T^\alpha_{\ \beta}T^\beta_{\ 0},$$
$$\{ T^0_{\ \rho},T^1_{\ 0}\}_q=qT^1_{\ \rho}-\lambda T^1_{\ \rho}
H_0, \quad
\{ T^0_{\ 1},T^\rho_{\ 0}\}_q=q^3T^\rho_{\ 1}-\lambda q^2 T^\rho_{\ 1}
H_0, \quad (\rho >1),$$
$$\eqalign{
\{T^0_{\ \rho},T^2_{\ 0}\}_q=qT^2_{\ \rho}-\lambda T^2_{\ \rho}H_0
-q\lambda T^2_{\ \rho}H_1 -\lambda q^3 T^1_{\ \rho}T^2_{\ 1}+\lambda^2
T^2_{\ \rho}H_1H_0+\lambda^2 q^2T^1_{\ \rho}T^2_{\ 1}H_0,\cr
(\rho >2),}$$
$$\eqalign{
\{T^0_{\ 2},T^2_{\ \rho}\}_q=q^3T^\rho_{\ 2}-\lambda q^2T^\rho_{\ 2}H_0
-\lambda q^3T^\rho_{\ 2}H_1 -\lambda q^3 T^1_{\ 2}T^\rho_{\ 1}+\lambda^2 q^2
T^\rho_{\ \rho}H_1H_0+\lambda^2 q^2T^1_{\ 2}T^\rho_{\ 1}H_0,\cr
(\rho >2),}$$
$$\{T^0_{\ 2},T^2_{\ 0}\}=H_0+qH_1-\lambda H_1H_0+q\lambda T^1_{\ 0}
T^0_{\ 1}-\lambda q^3T^1_{\ 2}T^2_{\ 1}+\lambda^2 q^2 T^1_{\ 2}T^2_{\ 1}
H_0,$$
$$\eqalign{
\{T^0_{\ 3},T^3_{\ 0}\}=& q^4H_0+q^3H_1+qH_2-\lambda q^2T^0_{\ 1}T^1_{\ 0}
-q\lambda T^0_{\ 2}T^2_{\ 0}-q\lambda H_0H_1-\lambda H_2 H_0 \cr
& -\lambda q^3T^2_{\ 3}T^3_{\ 2}-\lambda q^3T^1_{\ 3}T^3_{\ 1}-q\lambda H_1 H_2
-q^4\lambda^2T^1_{\ 2}T^2_{\ 1}+\lambda^2 q^3 T^2_{\ 3}T^3_{\ 2}H_1 \cr
& +\lambda^2 q^2 T^1_{\ 3}T^3_{\ 1}H_0+\lambda^2 q^2T^2_{\ 3}T^3_{\ 2}
H_0+\lambda^2H_1H_2H_0-\lambda^3 q^2T^2_{\ 3}T^3_{\ 2}H_1H_0\cr
& +\lambda^3q^3T^1_{\ 2}T^2_{\ 1}H_0,}$$
$$\eqalign{
\{T^0_{\ 4},T^4_{\ 0}\}=& -[T^1_{\ 4},T^4_{\ 1}]_{1/q^2}(q^4+\lambda H_0)
+q^6H_0-\lambda q^3T^0_{\ 1}T^1_{\ 0}-\lambda qT^0_{\ 2}T^2_{\ 0}
-\lambda qT^0_{\ 3}T^3_{\ 0} \cr
& +\lambda q^5T^1_{\ 3}T^3_{\ 1}+\lambda q^7T^1_{\ 2}T^2_{\ 1}
+\lambda^2 q^4 T^1_{\ 3}T^3_{\ 1}H_0+\lambda^2 q^6T^1_{\ 2}T^2_{\ 1}H_0,
}$$
where $I< \alpha< \beta <\gamma$, and for the other commutation relations
the generators satisfy the classical algebras, i.e., they commute or
anticommute depending on their grassman parity.

\end{document}